-------------------------------------------------------------------------------
        \documentstyle[preprint,tighten,aps,eqsecnum]{revtex}
        
\begin{document}

        \title{Vacuum Bubbles Nucleation and  Dark Matter Production
        through Gauge Symmetry Rearrangement.}
	\author{ S. Ansoldi\footnote{e-mail address:
         ansoldi@trieste.infn.it}}
        \address{ Dipartimento di Fisica Teorica dell'Universit\`a,\\
         Strada Costiera 11, 34014-Trieste\\
         Istituto Nazionale di Fisica Nucleare, Sezione di Trieste,\\
         Strada Costiera 11, 34014-Trieste, Italy,}

  	\author{ A. Aurilia\footnote{e-mail address:
         aaurilia@csupomona.edu}}
        \address{ Department of Physics, California State Polytechnic
        University\\ Pomona, CA 91768, USA,}

	\author{ E. Spallucci\footnote{e-mail address:
         spallucci@trieste.infn.it}}
       \address{\it Dipartimento di Fisica Teorica dell'Universit\`a,\\
         Strada Costiera 11, 34014-Trieste, Italy,\\
         Istituto Nazionale di Fisica Nucleare, Sezione di Trieste,\\
         Strada Costiera 11, 34014-Trieste, Italy,}

        \maketitle

        \begin{abstract}
	Modern particle physics and cosmology support the idea
	that a background of invisible material pervades the whole
	universe, and identify in the cosmic vacuum
 	the ultimate source of matter--energy, both seen and unseen.
 	Within the framework of the theory
	of fundamental relativistic membranes, we suggest a self--consistent,
	{\it vacuum energy--driven} mechanism for dark matter creation through
        gauge symmetry rearrangement.
	\end{abstract}

	\newpage
	\section{Introduction}
	\label{intro}

	The modern paradigms of physics are the standard Big Bang model of
cosmology and the standard
$SU(3)_{\rm C}\otimes SU(2)_{\rm L}\otimes U(1)_{\rm Y}$
model of the strong and electroweak interactions.
During the past two decades both models have been
refined with the addition of two key ingredients:
{\it inflation} on the cosmological side\cite{inflation} and {\it axions}
as pseudo Goldstone bosons associated with
the spontaneous breakdown of the Peccei-Quinn symmetry in particle
physics\cite{pq}.
Inflation requires the existence of dark matter and axions have long been
candidates for cold dark matter. A further refinement of the standard
models stems from a recent analysis of the cosmic microwave
background\cite{boomerang}
added to the data from high-redshift supernova
observations\cite{supernova}. Together, they seem to support the idea that
the universe is flat and is currently expanding at an accelerated
rate\footnote{At least the first acoustic peak in the power spectrum of
temperature fluctuations of the cosmic microwave background, as determined
by the Maxima and Boomerang observations, is best fit with $\Omega=1$
as required by inflation.}. As a result, dark matter {\it and} the
cosmological constant \cite{ostri}, or some form of dark energy, have become
the essential components of the new inflationary scenario\footnote{Those
revolutionary cosmological data were not available in 1991 when the authors
first suggested the possibility that the cosmological constant and dark
matter might be related \cite{essay}. This paper is based upon an essay
(unpublished)that was awarded {\it Honorable Mention} in the 1991 Gravity
Research Foundation competition.}.
In this paper we wish to suggest that those two
components are connected, in a rather fundamental way, by a new mechanism
of symmetry re-arrangement that {\it requires} the creation of dark matter.\\
Cold dark matter, in axionic form, could be detected in an experiment
capable of
probing masses in the range $10^{-6}\sim 10^{-3}$eV. Where does that mass
come from? The general consensus is that it comes from the quantum anomaly
which violates the chiral $U(1)_{\rm PQ}$ symmetry, thereby evading
Goldstone's theorem. However, the chiral anomaly is just one of at least
two possible loopholes
by which the existence of a Goldstone boson can be avoided. The second
loophole is the Higgs mechanism. In its conventional formulation, the Higgs
mechanism essentially converts a gauge field, i.e., a massless spin--1
field, into a massive vector field while preserving the value of the spin
as well as gauge invariance, thereby ensuring the renormalizability of the
theory. Even though this is the mechanism that generates the mass of all
known elementary particles within the Standard Model of particle physics,
it is clearly unsuitable to describe the cosmological situation envisaged
above, namely the conversion of the {\it constant} vacuum energy into
particles of matter. Thus we are led to ask: {\it How does one connect a
``constant energy background'' (non dynamical by definition) into material
particles that are
invisible but dynamical?} Our suggestion, in a nutshell, is as follows:
first, turn the cosmological constant into a {\it non dynamical gauge
field,} i.e., a gauge field with zero degrees of freedom; second, extract
from that gauge field a {\it massive spin--0 field} according to the time
honored procedure of symmetry breaking followed by restoration of gauge
invariance. As we shall see, the new ingredients of that old procedure are:
relativistic extended objects (membranes) and their gauge partners
(antisymmetric tensor gauge fields).\\
The topological nature of the new mechanism and its mathematical
formulation were discussed in a recent article in connection with the broad
issue of {\it electric--magnetic duality of p--branes}\cite{progress}. As
stated above, we are presently interested in applying the notion of
``topological symmetry breaking'' to the new inflationary scenario.
Accordingly, we have organized the paper as follows:\\
In Section(\ref{topsb}) we introduce the concept of topological
symmetry breaking. In Section(\ref{qed2}) we discuss the case of
``electrodynamics in two dimensions'', reinterpreted as ``bubble--dynamics
in two dimensions'', as the simplest framework
in which topological symmetry breaking can be implemented together with the
generation of mass.
In Section(\ref{dark}), the details of the new mechanism are illustrated in
four spacetime dimensions.
There we first outline the three main steps leading to the creation of
dark matter, and then discuss each step in a separate
subsection. Instrumental to the overall  mechanism is the false
vacuum decay rate through bubble nucleation from the vacuum which we
calculate in real spacetime and compare with the corresponding computation
in euclidean space.
Section(\ref{fine}) concludes the paper with a summary of our discussion and
a commentary on the applicability of the new mass generation mechanism to
the inflationary cosmological scenario.

	\section{Symmetry Breaking Revisited}
	\label{topsb}

``Topological symmetry breaking'' and the concomitant mechanism of mass
generation have never been discussed in the physics of {\it
point--particles} for the simple reason that the world history of a
material particle is usually assumed to have no boundary, that is, it is
usually assumed to be infinitely extended in time\footnote{In the
following, by ``boundary'' of an object (point--like or spatially extended)
we mean an extremal configuration (initial or final) of the world--history
of the object at a finite time, or, in other words, an extremal spacelike
section of the object's trajectory in spacetime. The case of an object with
a boundary that is {\it spatially} open can be treated along similar
lines\cite{as}\cite{aa}.}. Typically, in a world of classical point
particles described by a {\it local} field theory, reparametrization
invariance of the world--trajectory is tacitly assumed while gauge
invariance is {\it explicitly} broken only by introducing a mass term in an
otherwise invariant action. The asymmetry of the vacuum with respect to
some global transformation (spontaneous symmetry breaking) provides a
second possibility which, in turn, leads to the celebrated
Nambu--Goldstone--Higgs mechanism of mass generation. {\it The existence of
a boundary in the world history of an object provides an additional
possibility of symmetry breaking.}\\
For the sake of illustration, consider the familiar case of a charged
particle: the gauge invariance of the free Maxwell action may be broken
either by introducing a mass term into the action,
	 \begin{eqnarray}
	&&S\equiv S_m + S_A\\
	&&  S_m= -\mu_0 \, \int_0^\infty d\tau\, \sqrt{-{dx^\mu\over
	d\tau}\,{dx_\mu\over d\tau}}\\
	&&S_A=\int d^4x\,\left[-{1\over 4}\, F_{\mu\nu}\, F^{\mu\nu} +
	{m^2\over 2}\, A_\mu\, A^\mu
	-e\, J^\mu\, A_\mu\, \right]\label{snucl}
	\end{eqnarray}

or, more generally, by coupling the gauge potential $A_\mu(x)$ to a {\it
non--conserved} current $J^\mu(x)$, i.e., $\partial_\mu J^\mu(x)\ne 0$.
Thus, in either case: i) $m\ne 0$, or, ii)$m=0$, $\partial_\mu J^\mu\ne 0$,
gauge invariance 
is violated. {\it The  massless, non--conserved current case corresponds to a
classical point--like particle whose world--line $\Gamma_0$ has a free
end--point.} By definition, this represents a boundary condition that is
not explicitly encoded into the action. Therefore, to the extent that there
are no apparent symmetry violating terms in the action, we refer to this
case as ``topological symmetry breaking.'' For instance,

\begin{equation}
	\Gamma_0: x^\mu= x^\mu(\tau)\ ,\qquad 0\le \tau\le \infty
	\end{equation}
	with
\begin{equation}
	J^\mu(x)=\int_0^\infty\,  {dx^\mu\over d\tau}\,
	\delta^{4)}\left(\, x-x(\tau)\, \right) =\int_{\Gamma_0} dx^\mu\,
	\delta^{4)}\left(\, x-x(\tau)\, \right)
	\end{equation}
	represents a semi--infinite spacetime trajectory $\Gamma_0$ that
        originates at $x_0$ and then extends forever.
	 An extremal free end--point physically represents a ``singular''
        event in which a particle is either created or destroyed, so that the
        covariant conservation of the
	associated current is violated\footnote{
        A physical example of such a situation is the
	emission of an {\it alpha--particle} by a radioactive nucleus. Due
        to quantum tunneling, the
	particle suddenly disappears from within the nucleus ( its worldline
	comes to an end point) and reappears
	at a different point outside the parent nucleus. No {\it physical}
trajectory connects the two branches of the particle world line. Thus, from
the point of view of an external observer, the $\alpha$--particle world
line is semi--infinite: it originates from a
	point outside the nucleus at a given instant of time and then evolves
	independently of the parent nucleus.},

	\begin{equation}
	J(x)\equiv \partial_\mu \, J^\mu(x)=
	\partial_\mu \, \int_{\Gamma_0} d\, \left[\, \delta^{4)}
	\left(\, x-y(\tau)\, \right)\right]
	=\delta^{4)}\left(\, x-x_0\, \right)\ne 0\label{punto}
	\end{equation}

	Furthermore, under a gauge transformation of the action integral,
        the interaction term transforms as follows

	\begin{equation}
	\delta_\Lambda S^{int} =e\, \int_{x_0}^\infty dx^\mu\,
	\partial_\mu\, \Lambda= -\Lambda(x_0)
	\end{equation}
	assuming, as usual, that the gauge function vanishes at infinity.\\
In Section(\ref{dark}) we shall extend the above considerations to the case
of a relativistic bubble in $(3+1)$--dimensions. That is the natural setting
for discussing the new inflationary scenario. There we shall argue that the
corresponding classical action represents an {\it effective action} for the
quantum bubble nucleation process that takes place within the background
vacuum energy represented by the cosmological constant. The novelty here is
that the cosmological constant is disguised as a ``Maxwell field
strength''. Due to the presence of a boundary of the bubble trajectory in
spacetime, the process of nucleating an inflationary bubble \cite{bg},
\cite{noi4}, \cite{aussie}, \cite{noi5}
must be accompanied by the excitation of massive spinless particles. A
possible quantum formulation of
the same boundary mechanism using the path integral approach to the
dynamics of a generic p--brane in an arbitrary number of spacetime
dimensions is given in Ref.\cite{progress}.\\

	\section{ Two Dimensional Electrodynamics}
\label{qed2}
	``Electrodynamics in $(1+1)$ dimensions,'' also known in its early
quantum formulation as ``the Schwinger model'' \cite{schw}, means different
things to different people. Formally, the action (or Lagrangian) of the
model is the same as that of the familiar Maxwell electrodynamics in
$(3+1)$ dimensions, hence the name. The physical content, however, is
vastly different. This is because of the stringent kinematical constraints
that exist in $(1+1)$ dimensions: since there is no ``transversality'' in
one spatial dimension, the concept of spin is undefined, and the notion of
`` vector field'', massless or massive, is purely formal. Thus, there is no
radiation field associated with the Maxwell tensor. There is, however, the
same background vacuum energy and long range static interaction that we
shall discuss in the next section for the membrane theory in $(3+1)$
dimensions.
This is because in one spatial dimension a ``bubble'' degenerates into a
particle--anti particle pair, moving left and right respectively, and the
volume within the bubble is the linear distance between them\cite{acl}.
Indeed, the main reason for the following discussion
is to make it evident that those very kinematical constraints that
exist in $(1+1)$ dimensions are intertwined with the production of mass and
{\it can be induced just as well in $(3+1)$ dimensions} simply by
increasing the spatial dimensions of the object: from a 0--brane in $(1+1)$
dimensions to a 2--brane, or bubble, in $(3+1)$dimensions, indeed, to a
generic $p$--brane embedded in a target space with $p+2$--dimensions.
In other words, the familiar theory of electrodynamics in $(3+1)$
dimensions does not represent a {\it unique} generalization of the so
called ``electrodynamics in $(1+1)$ dimensions''. A more natural extension,
especially from a cosmological standpoint, is the theory of a relativistic
membrane coupled to a three index gauge potential. It is in the framework
of {\it bubble--dynamics}, regardless of the dimensionality of the target
space, that the cosmological constant drives the  creation of particles of
matter, and the engine of that process, at least at the classical
level, is the ``topological symmetry breaking'' due to the existence of a
boundary in the world history of the membrane.

	\subsection{Massless Phase }
	In the massless phase, the physical content of ``electrodynamics in
        $(1+1)$  dimensions, is encoded into the gauge invariant action :
	\begin{equation}
	S=\int d^2x\,\left[\, {1\over 4 }\, F^ {\mu\nu}\, F_ {\mu\nu}
	-{1\over 2 }\,
	F^{\mu\nu}\, \partial_ {\,[\,\mu} A_ {\nu\,]} -e\, J^\mu\, A_\mu\,
	\right]
	\label{essezero}
	\end{equation}

	so that the current density $J^\mu$, without further boundary
conditions, is divergenceless:
	$\partial_\mu\, J^\mu=0$. The first order formulation of the action
	is not mandatory but makes it clear that, in two dimensions, the
``Maxwell tensor'' is {\it assumed} to be the covariant curl of the gauge
potential, which is then treated as an independent variable.\\
	Thus, variation of the action with respect to the potential $A_\mu$
leads to
	the Maxwell equation

	\begin{equation}
	\partial_\mu\, F^ {\mu\nu}= e\, J^\nu
	\label{primo}\ .
	\end{equation}

	The general solution of Eq.(\ref{primo}) is the sum of the free
	equation solution ($e=0$), and a special solution of the inhomogeneous
	equation ($e\ne 0$). The complete equation can be formally
	solved by the Green function method. The final result is

	\begin{eqnarray}
	F^ {\mu\nu}(x)&=&\sqrt\Lambda \, \epsilon^ {\mu\nu} + e\,
	 \partial^{\,[\mu}\, {1\over \Box} \, J^{\nu\,]}\nonumber\\
	&=&\sqrt\Lambda \, \epsilon^ {\mu\nu} + e\, \int d^2y\,
	 \partial_x^ {\,[\mu}\, G(\, x-y\,) J^ {\nu\,]}(y)
	\end{eqnarray}
	 Inserting the above solution into the action (\ref{essezero}),
	 and neglecting surface terms, we obtain
	\begin{eqnarray}
	S&=&- {1\over 2} \int d^2x\,\left[\, \Lambda + e^2\,  J^ \nu\, { 1\over
	\Box }\, J_ \nu\,\right] \nonumber\\
	&=&- {1\over 2} \int d^2x\,\Lambda + e^2\,\int d^2x\, \int d^2y\,
	J^ \nu(x)\,G(\, x-y\,) \, J_ \nu(\, y\,)
	\label{2ds}
	\end{eqnarray}
	which we interpret as follows
	\begin{equation}
	S=- {1\over 2} \int d^2x\,\left[\, \hbox{ ``Cosmological Constant'' +
	Coulomb Potential  }\,\right] \ .
	\end{equation}
	 The first term represents a constant energy background,
	 or { \it cosmological term }, even though it can be
	``renormalized away'' in the absence of gravity.
	The second term in (\ref{2ds}) describes the long--range,
	``Coulomb interaction'' in two spacetime dimensions.
	In reality,  it represents the {\it linear confining potential }
	between point charges written in a manifestly covariant form. In such a
        covariant formulation, the existence of a boundary, even though not
        explicitly codified in the action (\ref{essezero}), introduces a
symmetry breaking condition since it implies that the world
        line of the ``charge'' has a free end-point through which the symmetry
        leaks out, so that $\partial_\mu\, J^\mu=J\ne 0$. In that case, gauge
        invariance is topologically broken and the current density is no longer
	divergence free. Under such circumstances, the field equation
        (\ref{primo}) needs to be modified since the left hand side is
        divergenceless, while the right hand side is not.

\subsection{Massive Phase}

The necessary remedy for the above inconsistency is the introduction of a
mass term in the action (\ref{essezero}). Paradoxically, the presence of
mass is also necessary in order to restore gauge invariance, albeit in an
extended form. As a matter of fact, the new action
\begin{equation}
	S=\int d^2x\,\left[\, {1\over 4 }\, F^ {\mu\nu}\,
	 F_ {\mu\nu} -{1\over 2}\,
	F^{\mu\nu}\,\partial_ {\,[\,\mu} A_ {\nu\,]} + {m^2\over 2 }\,
	A_\mu\, A^\mu -e\, J^\mu\, A_\mu\,\right]
	\label{esseuno}
	\end{equation}
reflects the fact that the original gauge invariance is not only
topologically broken, i.e., {\it implicitly} broken by the boundary, but
also {\it explicitly} broken by the
presence of a mass term. However, we argue that there is a subtle interplay
between
those two mechanisms of symmetry breaking, so that manifest gauge
invariance is actually restored. In order to further analyze the connection
between
the two mechanisms of symmetry breaking, it is convenient to separate the
divergenceless, boundary free current from the non conserved boundary
current. In two dimensions a generic vector can  be decomposed into the sum
of a ``hatted'', or divergence free component, and a ``tilded'', or curl
free
	 component:
	Thus, we write
	 \begin{equation}
	 A_\mu=  \widehat A_\mu +\widetilde A_\mu \ ,\quad:\quad
	  \partial_\mu\widehat A^\mu=0\ ,\quad
	  \partial_{[\,\mu}\widetilde A_{\nu\,]}=0\,
	\label{split}
	\end{equation}

	and a similar decomposition holds for the current,
	\begin{eqnarray}
	 &&J^\mu =\left(\, J^\mu- {\partial^\mu J\over \Box}\,\right) +
	 {\partial^\mu J\over \Box}
	 \equiv \widehat J^\mu + \widetilde J^\mu\ ,\quad:\quad
	\partial_\mu\widehat J^\mu =0\ , \quad\partial_\mu\widetilde
	J^\mu=J\ .
	 \end{eqnarray}

	In terms of this new set of fields and currents the action reads

	\begin{equation}
	S=\int d^2x\,\left[\, {1\over 4 }\, F^ {\mu\nu}\, F_ {\mu\nu}
	-{1\over 2 }\, F^{\mu\nu}\, \partial_ {\,[\,\mu}\widehat A_ {\nu\,]}
	+ {m^2\over 2 }\, \widehat A_\mu\,
	 \widehat A^\mu -e\, \widehat J^\mu\,  \widehat A_\mu+ {m^2\over
	 2 }\, \widetilde A_\mu\,\widetilde A^\mu -e\,
	 \widetilde J^\mu \, \widetilde A_\mu
	 \,\right]
	\label{essedue}
	\end{equation}

	 and we find two systems of decoupled field equations:
	 the divergence free vector field satisfies the  Proca--Maxwell
	 equation

	\begin{equation}
	\partial_\mu \, F^{\mu\nu} + m^2 \,\widehat A^\nu  = e\,
	 \widehat J^\nu
	 \label{f2}
	 \end{equation}

	while the curl free part must satisfy the constraint

	\begin{equation}
	 m^2 \, \partial_\nu \,\widetilde A^\nu
	 = e\, J
	\label{f3}
	\end{equation}

	or, equivalently

	\begin{equation}
	\widetilde A_\mu = {e\over  m^2  }\,\partial_\mu\,{1\over\Box}\, J
	\label{slong}
	\end{equation}

To the extent that the mass is linked to the divergence of the current, as
shown by the above equations, it is also a measure of the ``symmetry
leakage'' through the boundary. It is this connection between topological
and explicit symmetry breaking that leads us to ask: Is there a way of
restoring {\it manifest} gauge invariance in spite of the presence of a
mass term in the action?

\subsection{Massive, Gauge Invariant Phase}

The answer to the question raised in the previous subsection was suggested
by Stueckelberg a long time ago \cite{stuck}. The original Stueckelberg
proposal was to recover gauge invariance by introducing a compensating scalar
field $\theta$ so that the resulting action

	\begin{equation}
	S_A=\int d^2x\, \left[\,{1\over 4}\, F_{\mu\nu}\,
	F^{\mu\nu} -{1\over 2}\, F^{\mu\nu}\,
	\partial_{\,[\,\mu} \, A_{\nu\,]}
	+{m^2\over 2}\, \left(\, A_\mu +{1\over m }\,
	\partial_\mu \theta\, \right)^2 -
	 e\, J^\mu \,\left(\, A_\mu +{1\over m }\,
	 \partial_\mu\, \theta\, \right)\, \right]
        \label{stuckm}
	\end{equation}

	is invariant under the extended gauge transformation

	\begin{eqnarray}
	 A_\mu &&\longrightarrow A_\mu^\prime = A_\mu + \partial_\mu\,
	 \lambda \label{ext1}\\
	\theta &&\longrightarrow \theta^\prime=\theta-m\, \lambda\ .
	\label{ext2}
	\end{eqnarray}
In this case, the vector field equation

	\begin{equation}
	\partial_\mu\, F^{\mu\nu}+m^2\,
	\left(\, A^\nu + {1\over m }\, \partial^\nu\, \theta\, \right) =
	 e\, J^\nu
	\end{equation}
	 is self--consistent because of the $theta$--field equation
	\begin{equation}
	 m^2\,\partial_\nu\,\left(\, A^\nu + {1\over m }\,\partial^\nu\,
	 \theta\, \right) = e\,\partial_\nu\, J^\nu\ .
        \label{thetaeq}
	\end{equation}

	Evidently, the role of the constraint (\ref{thetaeq}) is to combine
        the $\widetilde A_\mu$
	component of the vector potential with the compensator field in
such a way that symmetry is restored with respect to the extended gauge
transformation. In our geometric interpretation, this is equivalent to
``closing the world history'' by compensating for the leakage of symmetry
through the boundary. In this sense, the generation of mass is the
consequence of ``mixing'' two gauge fields, namely the $\widetilde A_\mu$
	 component of the vector potential with the $\theta$--field. As a
matter of fact, Eq. (\ref{thetaeq}) determines the {\it mixed, gauge invariant}
 field to be

	\begin{equation}
	\widetilde A_\mu +{1\over m }\, \partial_\mu\, \theta = {e\over m^2}\,
	\partial_\mu \, {1\over\Box}\, J
	\label{longii}
	\end{equation}

	 Once the above equation (\ref{longii}) is inserted into the action
        (\ref{stuckm})
	 we obtain
	  \begin{equation}
	S=\int d^2 x\, \left[\, {1\over 4}\, F_{\mu\nu}\, F^{\mu\nu}  -{1\over
	2}\, F^{\mu\nu}\, \partial_{\,[\,\mu} \, A_{\nu\,]}
	+{m^2\over 2} \,\widehat A_\mu^2 - e\,\widehat J^\mu\,
	\widehat A_\mu
	-{e^2\over 2m^2 }\,J\, {1\over \Box}\, J\,\right]\label{stuckmm}
	\end{equation}

	which represents an ``effective action'' for the
	only physical degree of freedom represented by $\widehat A_\mu$.

	\section{The Three Stepping Stones of `Dark Matter' Production:
	\\Formulation of the Mechanism}
	\label{dark}
 In order to place our previous discussion in the right perspective and
partly to justify the more technical approach in the following subsections,
let us consider the inflationary idea that the early phase of the exponential
expansion of the universe inflated a microscopic volume of space
to a size much larger that the presently observable part of the universe;
this idea can be formulated
within the framework of General Relativity as a special case of ``Classical
Bubble Dynamics'' (CBD), i.e., the study of the evolution of a vacuum
bubble in the presence of gravity \cite{bg}. In our own formulation of CBD,
inflation is driven by a gauge field $A_{\mu\nu\rho}(x)$
which {\it is equivalent to a cosmological constant} \cite{noi4}, and the
boundary effects in CBD, completely similar to those discussed in the
previous section, constitute the precise mechanism which extracts dark
matter from the self--energy of $A_{\mu\nu\rho}$.\\
In short, how does that process take place? The following properties of
$A_{\mu\nu\rho}$ constitute the crux of the boundary mechanism in the
inflation--axion scenario:\\
\\
a) When massless,
 $A_{\mu\nu\rho}$ represents `` dark stuff '' by definition, since in
(3+1)--dimensions $A_{\mu\nu\rho}$ does not possess radiative degrees of
freedom. In fact, the field strength
$ F_{\mu\nu\rho\sigma}\equiv\nabla_{[\,\mu}\, A_{\nu\rho\sigma]}=
\partial_{[\,\mu}\, A_{\nu\rho\sigma\, ]}$, as a solution of the
classical field equation, is simply a constant disguised as a
gauge field. This property, even though peculiar, is not new in
field theory: it is shared by all $d$--potential forms in
$(d+1)$--spacetime dimensions.
For instance in two dimensions,
$F_{\mu\nu}=\partial_{[\,\mu}A_{\nu]}=\epsilon_{\mu\nu}\,\Lambda $ while in
four dimensions, $F_{\mu\nu\rho\sigma}=\epsilon_{\mu\nu\rho\sigma}\, f$,
and $f$
represents a constant background field in both cases by virtue of the field
equations. What is then the meaning of ``$f$''?
{\it As a gauge field, $A_{\mu\nu\rho}$ is
endowed with an energy momentum tensor and thus it couples to gravity
\cite{acl}: the resulting equations are Einstein's equations with the
cosmological term $\Lambda=4\,\pi\, G\, f^2$.}
For this reason we call $A_{\mu\nu\rho}$ the ``{\bf cosmological field}''.
This alternative interpretation of the cosmological constant can be traced
back to Ref.(\cite{acl}) and its application to the inflationary scenario
in Ref.(\cite{noi4}); it will be discused in more detail in the following
subsection.\\
\\
b) If the cosmological field acquires a mass, then it describes {\it
massive pseudoscalar} particles, in contrast with the usual Higgs
mechanism. Indeed, in the massive case the free field equation for
$A_{\mu\nu\rho}$

	\begin{eqnarray}
	&&\partial_\lambda \partial^{\,[\,\lambda}
	A^{\mu\nu\rho\,]}+m^2
        A^{\mu\nu\rho}=0\ , \quad \Longrightarrow \partial_\mu \,
        A^{\mu\nu\rho}=0
        \label{unobis}
        \end{eqnarray}

imposes the divergence free constraint on the four
components of $A_{\mu\nu\rho}$ leaving only one {\it propagating}
degree of freedom. {\it In other words, the introduction of a mass term
``excites'' a dynamical (pseudoscalar) particle of matter out of the
cosmological energy background.}\\
\\
c) Evidently, the transition from case a) (massless, non dynamical field)
to case b) (massive propagating particles) requires a physical mechanism
for its enactment. {\it Here is where the idea of topological symmetry
breaking and the concomitant rearrangement of gauge symmetry come into
play.} We hasten to say here, and expand our discussion in the following
subsection, that the cosmological field $A_{\mu\nu\rho}$ does not interact
directly with
the ordinary matter fields that represent point--like particles. Rather,
$A_{\mu\nu\rho}$ is
the ``gauge partner'' of relativistic closed membranes,
or bubbles, in the sense that it mediates the interaction between surface
elements according to the same general principle of gauge invariance
which dictates the coupling of point charges to vector gauge bosons, or the
coupling of Kalb--Ramond potentials to elementary string--like objects.
Clearly, this type of coupling to relativistic membranes as fundamental
extended objects, is a crucial assumption of the whole mechanism of mass
generation advocated in this paper.

\subsection{ Massless Field, Closed Membrane and Vacuum Energy Density}

In order to implement the three properties a), b), c) discussed above, we
start from the action functional

	\begin{equation}
	S=\int d^4x\,\left({ 1\over 2\cdot 4!}\, F_ {\lambda\mu\nu\rho}\,
	F^ {\lambda\mu\nu\rho} -{ 1\over  4!}\, F^ {\lambda\mu\nu\rho}\,
	\partial_{\,[\lambda}\, A_{\mu\nu\rho\,]}
	- { g\over 3!}\, A_{\mu\nu\rho}\, J^{\mu\nu\rho}
	\,\right)-\mu^3\,\int_M d^3\sigma\, \sqrt{-\gamma }\ .
       \label{cbd}
        \end{equation}

This is a straightforward, but non--trivial, formal extension of the action
for the electrodynamics of point charges, or the Kalb--Ramond  action of
``string--dynamics.'' More to the point, from our discussion
inSection(\ref{qed2}), {\it it represents a direct generalization to $(3+1)$
dimensions of the same two--dimensional``electrodynamics'' action}
\cite{acl}  {\it in a ``$\sigma$--model'' inspired  formulation where a 
fundamental extended object is coupled to its massless excitations.} In order to
 keep our discussion as transparent as possible, we consider only an elementary,
or {\it structureless} membrane with
vanishing width, interacting with a single massless mode represented by
$A_{\mu\nu\rho}$. Thus,

	\begin{equation}
	 J^{\mu\nu\rho}(\, x\, )=\int \,\delta^{ 4)}
	 \left[\, x- Y(\sigma)\,\right]\,
	 dY^\mu \wedge dY^\nu\wedge dY^\rho
	 \label{mc}
	 \end{equation}
	represents the current density associated with the world history of
        the membrane. More complex models,
in which the membrane is some sort of collective excitation of an underlying
field theory, while intriguing, are affected by highly non--trivial technical
problems, such as  renormalization \cite{nee},\cite{as} and bosonization
\footnote{
Introducing fermionic degrees of freedom enables one to establish a
correspondence between
bosonic and fermionic variables, that is,  $J^{\mu\nu\rho}\leftrightarrow
\epsilon^{\mu\nu\rho\sigma} \bar\psi\,\gamma^5\, \gamma_\sigma\,\psi$. }.
That approach, while conceivable in principle, is orthogonal to ours: here,
we {\it assume} that the membranes under consideration are elementary
geometric objects of a fundamental nature, on the same footing as points,
strings and other p--branes that constitute the very fabric of quantum
spacetime\cite{essay2}. This principle of {\it geometric democracy,} is
reflected in the action functional (\ref{cbd}) by our choice of
the Nambu--Goto--Dirac action for a relativistic bubble in which $\gamma$
stands for the determinant of the induced metric
	 \begin{equation}
	 \gamma\equiv det\left(\,\partial_m Y^\mu \,\partial_n Y_\mu\,\right)
	  \end{equation}
	  and $\mu^3$ represents the bubble surface tension.
	  Notwithstanding the apparent simplicty of our model, we shall
	  see in the next subsection that it is possible to reproduce the
	  correct false vacuum decay rate, without resorting to solitonic
	  computational techniques\footnote{It seems to us that this result
	may well be
	  a consequence of a {\it duality} between membranes as solitonic
	  solutions of an underlying field theory and membranes as fundamental
	  objects (for a comprehensive review, see Ref.\cite{dual}).}.
	   \\
	Gauge invariance of the action (\ref{cbd}) is guaranteed whenever
the bubble embedding equations
	$x^\mu=Y^\mu(\,\sigma\,)$ parametrize a world history without
	boundary, so that

	\begin{eqnarray}
	&& \delta A_ {\mu\nu\rho} =\partial_{[\, \mu}\, \Lambda_{\nu\rho\,]}
	\\
	&& \delta S_\Lambda = 0 \quad\longleftrightarrow \quad
	\partial_\mu\, J^{\mu\nu\rho}=0\label{div0}
	\end{eqnarray}

	The divergence free condition for the membrane current is the formal
	translation of the no--boundary condition.
	Essentially, it restricts the world history of the membrane to be
	\cite{teit}:\\
	i) { \it spatially closed; }\\
	ii) either  infinitely extended in the timelike direction ({\it
	 eternal membrane ,}) or, compact without boundary ({\it virtual
	 membrane.}) \\
	Consequently, the cosmological field $A_ {\mu\nu\rho}$  couples in
	 a gauge invariant way {\bf only} to bubbles whose history extends
	 from the remote past to the infinite future,
	 or to objects that start as a point in the vacuum, expand to a
	maximum spatial volume and then recollapse to a point in the vacuum.
	In such a case, variation of the action with respect to
	$A_{\mu\nu\rho}$ leads to Maxwell's equation
	 \begin{equation}
	  \partial_\mu\, F^ {\mu\nu\rho\sigma}= g\, J^ {  \nu\rho\sigma}
	\label{max1}
	\end{equation}

	The general solution of (\ref{max1}) is the sum of the free
	equation solution ($g=0$),  and a special solution of the
	inhomogeneous equation ($g\ne 0$). The complete formal solution is
	found by inverting the field equation according to the Green function
	method: taking into account that the Maxwell tensor is proportional
	to the { \it epsilon--tensor}, we find

	\begin{equation}
	F^{\mu\nu\rho\sigma}= \epsilon ^ {\mu\nu\rho\sigma}\,
	 \sqrt\Lambda + g\,\partial^ {\,[\mu}\,{1\over\Box}\,
	  J^{\nu\rho\sigma\,]}\ .
	\label{classol}
	\end{equation}

	Inserting the above solution back into the action (\ref{cbd}), one
        finds apart from the Nambu--Goto--Dirac term

	\begin{equation}
	S= - { 1\over 2}\int d^4x\,\left(\, \Lambda + { g^2\over  3!}\,
	J^{\nu\rho\sigma}\, { 1\over \Box }\,  J_ {\nu\rho\sigma}\,\right)\ .
	\label{cbdshort}
	\end{equation}

	Exactly as in the $(1+1)$ dimensional case, we interpret the above
expression as follows

	\begin{equation}
	S=- {1\over 2} \int d^4x\,\left[\, \hbox{ ``Cosmological Constant'' +
	``Coulomb Potential''  }\,\right] + \hbox{ extra surface terms}
	\end{equation}

	Indeed, as anticipated at the beginning of this section (property (a)),
the first term in Eq.(\ref{cbdshort}) is a solution of the {\it free}
Maxwell equation and represents a constant energy density background. As a
free field, that is, in the
absence of gravity and any other interaction, that constant term
can be ``renormalized away'' since it cannot be distinguished from the
vacuum. However, it is equivalent to a cosmological term when gravity is
switched on\cite{acl},\cite{noi5}.

	It may not be immediately evident
	that, even in the presence of a coordinate dependent metric
	$g_{\mu\nu}(x)$, the homogeneous solution of Eq.(\ref{primo}) still
represents
	a constant background energy density. Phrased differently, it might
	appear that ``there is no longer a constant rank--$4$ tensor
available
	to equate $F$ to''.\\
	The loophole is in the covariant form of Eq.(\ref{primo}). Since we
	are considering the homogeneous solution, we may as well
switch--off the coupling to the
	current, so that:
	\begin{equation}
	\nabla_\mu\, F^{\mu\nu\rho\sigma}=0\label{2dscov}\ .
	\end{equation}
	Here, $\nabla_\mu$ represents the covariant derivative compatible with
	the Riemannian metric $g_{\mu\nu}(x)$, i.e., the connection
	is chosen to be the Christoffel symbol.  In four dimensions there
	is only one generally covariant and totally anti--symmetric tensor,
	namely, the covariant Levi--Civita {\it tensor}:
	\begin{equation}
	\varepsilon^{\mu\nu\rho\sigma}(x)\equiv {1\over \sqrt{-g(x)}}\,
	\epsilon^{\mu\nu\rho\sigma}
	\end{equation}
	where,  $g(x)\equiv det\, g_{\mu\nu}(x)$ and
	$\epsilon^{\mu\nu\rho\sigma}$ is the constant Levi--Civita tensor
	density. Thus, Eq.(\ref{2dscov}) may be solved by the ansatz

	\begin{equation}
	 F^{\mu\nu\rho\sigma}\equiv {1\over \sqrt{-g(x)}}\,
	\epsilon^{\mu\nu\rho\sigma}\, F(x) \label{trial}
	\end{equation}

	where $F(x)$ is a {\it scalar } function to be determined by the
	field equations.
	The metric tensor $g_{\mu\nu}(x)$ and its determinant are both
	covariantly constant with respect to the  Christoffel covariant
	derivative. Thus, the $\varepsilon(x)$ tensor has vanishing covariant
	derivative. By inserting the trial solution (\ref{trial}) in
	Eq.(\ref{2dscov}), one sees that the derivative operator bypasses
	the $\varepsilon(x)$ tensor and applies directly to the scalar
function F(x):

	\begin{equation}
	\varepsilon^{\mu\nu\rho\sigma}(x)\partial_\mu\, F(x)=0\label{3dscov}
	\end{equation}

	Thus, the solution of Eq.(\ref{3dscov}) is again

	 \begin{equation}
	F(x)=\, const.\equiv \sqrt\Lambda\label{4dscov}
	 \end{equation}

	To conclude the proof that $\Lambda$ represents a genuine
cosmological constant,
	we need to compute the value of the classical action. This can be
	done by using the following property of the $\varepsilon(x)$ tensor:

	\begin{equation}
	\varepsilon_{\mu\nu\rho\sigma}(x)\equiv
	\sqrt{-g(x)}\, \epsilon_{\mu\nu\rho\sigma}\ ,\qquad \longrightarrow
    \varepsilon_{\mu\nu\rho\sigma}(x)\,\varepsilon^{\mu\nu\rho\sigma}(x)=-4!
	\end{equation}

	 Thus,

	 \begin{equation}
	 -{ 1\over 2\cdot 4!}\int d^4x\,\sqrt{-g}\, F_ {\lambda\mu\nu\rho}\,
	F^ {\lambda\mu\nu\rho}\longrightarrow
	\int d^4x\,\sqrt{-g}{ \Lambda\over 2}\ .
	 \end{equation}

	Having clarified the physical meaning of the integration constant
	$\Lambda$, let us consider the second term in the action
	(\ref{cbdshort}).  Apparently, it describes a
       long--range, ``Coulomb interaction'' between the bubble surface
elements.
	In reality, it represents the bubble {\it volume energy density}
	written in a manifestly covariant form. In fact, we can re--arrange
	that ``Coulomb term'' as follows. From the
	definition (\ref{mc}) and the condition (\ref{div0}), we deduce that
	\begin{eqnarray}
	J^ {\nu\rho\sigma}(x) &=& \partial_\mu\, K^{\mu\nu\rho\sigma}(x)\\
	&=&\partial_\mu \int d^4\xi \, \delta^{4)}\left[\, x -Z(\xi)\,\right]
	dZ^\mu\wedge dZ^\nu\wedge dZ^ \rho\wedge dZ^\sigma\ .
	\end{eqnarray}
	However, in four dimensions
	\begin{equation}
	K^{\mu\nu\rho\sigma}(x)=\epsilon ^ {\mu\nu\rho\sigma}\int d^4x
        \,\delta^{4)}\left[\, x -Z(\xi)\,\right]\equiv
	\epsilon ^ {\mu\nu\rho\sigma}\Theta(x)
	\end{equation}

	where $\Theta(x)$ is referred to as the { \it characteristic
        function} of
	the spacetime open sub--manifold  bounded by the membrane.
	Thus, the Coulomb term can be rewritten in terms of $\Theta(x)$

	\begin{eqnarray}
	{ g^2\over  3!}\,J^ {\nu\rho\sigma}\, { 1\over \Box }\,
	J_ {\nu\rho\sigma}&=& { g^2\over  3!}\partial_\mu \,
	K^{\mu\nu\rho\sigma}
	{1\over \Box }\,\partial^\tau\, K_{\tau\nu\rho\sigma}\nonumber\\
	&=& { g^2\over  3!}\,\partial_\mu\,
\epsilon^{\mu\nu\rho\sigma}\Theta(x)
	\,{ 1\over \Box }\, \partial^\tau \epsilon_{\tau\nu\rho\sigma}\Theta(x)
	\nonumber\\
	&=& -g^2\partial_\mu\, \Theta\, { 1\over \Box }\, \partial^\mu\,\Theta
	\nonumber\\
	&=& g^2 \Theta(x)
	\end{eqnarray}
	where we have made use of the formal identity $\Theta^2(x)\equiv
	\Theta(x)$ and discarded a total divergence.
	Thus, the classical solution (\ref{classol}) and the action
(\ref{cbd}) show that the cosmological field $A_ {\mu\nu\rho}$
	does not describe the propagation of material particles; rather, it
represents a constant energy density background with two different
	values inside and outside the membrane. Indeed, using the previous
        result, one may calculate the value of
	the classical action corresponding to the solution (\ref{classol}),

	\begin{equation}
	S =-{ 1\over 2}\int d^4x\,\left[\, \Lambda + g^2 \,\Theta(x)\,
	\right] -\mu^3\int_M d^3\sigma\, \sqrt{-\gamma }\ .
	\label{13}
	\end{equation}

	Once again, we note that in the absence of gravity one is at
	liberty to choose the ``zero'' of the
	energy density scale, and thus measure the energy density with respect
	to the constant background represented by $\Lambda$. With that
	observation in mind, the classical action turns out to be a
	{ \it pure volume} term, as announced:

	\begin{eqnarray}
	S_\Lambda +\mu^3\int_M
	d^3\sigma\, \sqrt{-\gamma }
	&\equiv& S(\, \Lambda; g\,)- S(\,\Lambda; g=0\,) =
	{ g^2\over 2}\, \int d^4x\, \Theta(x)  \nonumber\\
	&=& { g^2\over 2}\, \int d^4x\, \int_B d^4\sigma\,
	\delta^ {4)}\left[\, x - Y(\sigma)\,\right]
	\label{svol}
	\end{eqnarray}

\subsection{Nucleation Rate, Symmetry breaking and Mass}
\label{clmass}

With the results of the previous subsection in hands, we can finally relate
the nucleation rate of vacuum
	bubbles with the idea of topological symmetry breaking and mass
generation. A common procedure for computing the nucleation rate of vacuum
bubbles amounts, in our present formulation, to analytically continue the
action (\ref{svol}) to imaginary time \footnote{$\epsilon^{\mu\nu\rho\sigma}
\rightarrow
	i\varepsilon^{\mu\nu\rho\sigma}$ under Wick rotation. Thus,
	\begin{equation}
	  \partial_\mu\, K^ {\mu\nu\rho\sigma}\, { 1\over\Box}\,
	    \partial^\tau\, K_{\tau\nu\rho\sigma}\rightarrow
	+g^2 \partial_\mu\,\Theta(x)\,{ 1\over\Box}\,\partial_\mu\,\Theta(x)\ .
	\end{equation} }

	\begin{eqnarray}
        S_E(\, F\,; g^2\, )=
        {1\over 2}\int d^4x \left[\, \Lambda - g^2\,\Theta_B(x)\,\right]
	-\mu^3\, \int d^p\sigma \sqrt{-\gamma}
        \end{eqnarray}
	A semi--classical estimate for the nucleation rate of a {\it  spherical
	bag } of radius $R$ can be obtained through a saddle point estimate
	of $S_E$ :

	\begin{eqnarray}
	e^{-\Gamma}&\simeq& e^{ -[ \, S(\,\Lambda \,; g\, )- S(\,\Lambda \,;
	0\,)\, ]}
	\equiv e^{- S(R) }\nonumber\\
	&=& \exp\left[-
	\left({\pi^2\over 4}g^2 R^4 -2\pi^2\mu^3 R^3\right)\Big\vert_{R=R_0}
	\,\right]
	\end{eqnarray}
	where the {\it nucleation radius} $R_0$ is a stationary point
	\begin{equation}
	\left({\partial   S(R)\over \partial R}\right)_{R=R_0}
	\quad \longrightarrow \quad R_0={6\mu^3\over g^2}\ .
	\end{equation}
	Then, one finds
	\begin{equation}
	e^{-B}=\exp\left(-{\pi^2\over 2}\mu^3 R_0^3\right)
	\end{equation}

	which is the original Coleman De Luccia result for the false
	vacuum decay rate\cite{coleman}.\\
	Apart from confirming the validity of our approach against a well
tested calculation, the euclidean description of vacuum decay shows how a
	vacuum bubble may materialize in Minkowski spacetime as a
	spacelike domain at a finite time. This is precisely the
       ``extremal'', i.e. initial boundary of the membrane
	world manifold, and leads us to conclude, by the argument of the
previous section, that the current associated with the bubble nucleation
process cannot be divergence free. Thus, a Minkowskian description of the
bubble nucleation
	process seems to be in conflict with the requirement of gauge
	invariance.\\
	Against this background, we wish to show that a consistent
	description, in real spacetime, of the (quantum) nucleation
process, can be achieved by restoring the gauge invariance of the original
action. However, restoring gauge invariance is tantamount to ``closing''
the world history of the membrane, so that no boundary exists. There are
essentially two ways to achieve this: in the euclidean
	formulation one bypasses the problem  by ``closing the free
        boundary in imaginary time'', so that the resulting euclidean world
manifold is again without boundary. Somewhat paradoxically, the alternative
procedure in real spacetime is to include in the action an additional
source of symmetry violation in the form of a mass term for the
cosmological field. We hasten to emphasize, before proceeding further, that
the inclusion of a mass term is not a matter of choice. As we have seen
in $(1+1)$ dimensions, it
is actually dictated by the self--consistency of the field equations.
There, we have shown how the explicit symmetry breaking due to the presence
of mass
and the topological symmetry breaking due to the presence of a boundary
actually conspire to produce an action which is gauge invariant, albeit in
an extended form. Thus, in the last analysis, {\it it is the
self--consistency of the theory that forces upon us the introduction of
a massive particle.}\\
\\
Consider the coupling of the cosmological field $A_{\mu\nu\rho}(x)$ to a
quantum mechanically nucleated relativistic membrane. According to the
discussion in the previous section, the history of such
	an object is spatially closed, but {\it  only semi--infinite} along
	the timelike direction because the membrane comes into existence
	at a finite instant of time. The nucleation event provides a spacelike
 boundary that consists of a two--surface where symmetry ``leaks
out'' and gauge invariance is broken. Therefore, the {\it apparently} gauge
invariant action

	\begin{equation}
        S_0=\int d^4x\,\left(\, {1\over 2\cdot 4!}\,F_{\lambda\mu\nu\rho}\,
        F^{\lambda\mu\nu\rho}
        -{ 1\over  4!}\, F^ { \lambda\mu\nu\rho}\,
        \partial_{\,[\lambda } A_ {\mu\nu\rho\,]}
        -{g\over 3!}\, J^{\mu\nu\rho}\, A_{\mu\nu\rho} \,\right)
        \label{essezeroo}
	\end{equation}

	leads to field equations

	\begin{equation}
	\partial_\lambda\, F^{\lambda\mu\nu\rho}= g\, J^{\mu\nu\rho}(x)
	\label{bleah}
	\end{equation}

that are inconsistent. This is because the l.h.s. of Eq.(\ref{bleah})is
divergence free everywhere due to the antisymmetry of the Maxwell tensor,
whereas the membrane current
	is divergenceless everywhere except at the nucleation event where

	\begin{equation}
	\partial_\mu\, J^{\mu\nu\rho}=j^{\nu\rho}\ne 0\ .
	\end{equation}

	Here $j^{\nu\rho}$ represents the { \it boundary current} localized
	on the initial two--surface. For a Minkowskian observer the membrane is
	created { \it ex nihilo, }and its current suddenly jumps from
	zero to a non--vanishing value. Therefore, it cannot be
	``conserved'' and
	the amount of (topological) symmetry breaking is taken into account
	by $j^{\nu\rho}$. Thus, what we learn from Eq.(\ref{bleah}) is that
	{ \it the massless cosmological field {\bf cannot couple} to the
	current of a relativistic membrane which is nucleated from the
	vacuum}. \\
	In order to write down a self--consistent model for interacting
	semi--infinite world histories, the coupling must involve a
	{ \it massive } tensor field:

	\begin{equation}
        S=\int d^4x\, \left(\,{1\over 2\cdot 4!}\,F_{\lambda\mu\nu\rho}\,
        F^{\lambda\mu\nu\rho} -{ 1\over  4!}\, F^ { \lambda\mu\nu\rho}\,
        \partial_{\,[\lambda } A_ {\mu\nu\rho\,]}
        +{ m^2\over 2\cdot 3!}\, A_{\mu\nu\rho}\, A^{\mu\nu\rho}
        -{g\over 3!}\, J^{\mu\nu\rho}\, A_{\mu\nu\rho} \,\right)\ .
        \label{massmem}
	\end{equation}

Inspection of the above action tells us that the physical spectrum consists
of massive spin--0 particles, in agreement with property b) listed in the
previous section. However, in order to extract the full physical content of
the system (\ref{massmem}) we can proceed as follows:\\
from the above action we derive the field equations

	\begin{eqnarray}
	&& \partial_\lambda\, F^{\lambda\mu\nu\rho} +m^2\, A^{\mu\nu\rho}=
	    g\, J^{\mu\nu\rho}\label{tmuno}\\
	&&  \partial_\mu A^{\mu\nu\rho}=  {g\over m^2   }\, j^{\nu\rho}\ .
	\label{tmdue}\\
	\end{eqnarray}

	Next, we use the identity

	\begin{equation}
J^{\mu\nu\rho}=\left(\, J^{\mu\nu\rho}-\partial^{\, [\,\mu}\, {1\over \Box}
	\,j^{\nu\rho\,]}\,\right) +\partial^{\, [\,\mu}\,{1\over \Box}\,
	j^{\nu\rho\,]}
	\equiv \widehat J^{\mu\nu\rho}+  \widetilde J^{\mu\nu\rho}\\
	\end{equation}

in order to split the current into two parts
\begin{equation}
\partial_\mu\widehat J^{\mu\nu\rho}=0\ ,\qquad \partial_\mu\widetilde
	J^{\mu\nu\rho}=j^{\nu\rho}\ .
	\end{equation}

Evidently, $\widehat J^{\mu\nu\rho}(x)$ is the divergenceless, boundary
free current, while $\widetilde J^{\mu\nu\rho}(x)$ represents the
	pure boundary current.\\
	In a similar fashion we decompose the tensor gauge field into the sum
	of a { \it divergence free} and a { \it curl free} part:
	\begin{eqnarray}
	&& A^{\mu\nu\rho}\equiv \widehat A^{\mu\nu\rho} + \widetilde
	A^{\mu\nu\rho}\\
	&& \partial_\mu\,\widehat A^{\mu\nu\rho}=0\ ,\quad
	\partial_{[\,\lambda}\, \widetilde A_{\mu\nu\rho\,]}=0\ .
	\end{eqnarray}

	Consequently, the ``Proca--Maxwell'' equations (\ref{tmuno}),
	(\ref{tmdue}) split into the following set

	\begin{eqnarray}
	&& \partial_\lambda\, F^{\lambda\mu\nu\rho} +m^2\,\left(\,
	\widehat A^{\mu\nu\rho} + \widetilde A^{\mu\nu\rho}\,\right)  =
	    g\,\left(\, \widehat J^{\mu\nu\rho} + \widetilde
	    J^{\mu\nu\rho}\, \right)
	\label{vmunobis}\\
	&&  \partial_\mu \,\widetilde A^{\mu\nu\rho}=
	{g\over m^2   }\, j^{\nu\rho}\ .
	\label{vmduebis}
	\end{eqnarray}

	From eq.(\ref{vmduebis}) we obtain

	\begin{equation}
	\widetilde A^{\mu\nu\rho}=
	{g\over  m^2  }\,\partial^{\, [\,\mu }\, {1\over \Box}\,j^{\nu\rho\,]}
	\ .
	\label{tlong}
	\end{equation}

	Then, Eq.(\ref{vmunobis}) becomes

	\begin{equation}
	\partial_\lambda\, F^{\lambda\mu\nu\rho} +m^2\,\widehat A^{\mu\nu\rho}
	= g\, \widehat  J^{\mu\nu\rho}
	\end{equation}

	which gives

	\begin{eqnarray}
	F^{\mu\nu\rho\sigma}&=& \sqrt\Lambda\,\epsilon^{\mu\nu\rho\sigma}\,
	+ g^2\,  \partial^{\, [\,\mu }\,{1\over \Box}\,
	\widehat J^{\nu\rho\sigma\,]} -m^2\,\partial^{\, [\,\mu}\,
	{1\over \Box}\,\widehat A^{\nu\rho\sigma\,]}
	\nonumber\\
	&\equiv & F^{\mu\nu\rho\sigma}_0 -m^2\,
	\partial^{\, [\,\mu}\,{1\over\Box}\, {\widehat A}^{\nu\rho\sigma\,]}
	\label{massivesol}
	\end{eqnarray}

	where $F^{\mu\nu\rho\sigma}_0$ represents the solution of Maxwell's
	equation

	\begin{equation}
	\partial_\mu\, F^{\mu\nu\rho\sigma}_0= g\, \widehat
	J^{\nu\rho\sigma}
	\end{equation}

	obtained in the previous massless case.

	Substituting the solution (\ref{massivesol}) into the action
(\ref{massmem}), and following step by step the same procedure outlined in
	the previous subsection for the massless case, we obtain the
	corresponding result for the massive cosmological field

        \begin{eqnarray}
        S =  \int d^4 x &&\, \left[\,  {1\over 2\cdot 4! }\,
        F_ {0\,\,\mu\nu\rho\sigma}\,F_0^{\mu\nu\rho\sigma}
        -{m^2\over 2\cdot 3!}\,
        \widehat A^{\mu\nu\rho}\, \left(\, {\Box +  m^2\over \Box}\,\right)
	\widehat A_{\mu\nu\rho}
	+ {g\,m^2\over 3!}\,
	\widehat J^{\mu\nu\rho}\, {1 \over\Box}\,\widehat A_{\mu\nu\rho}
	\right.\nonumber\\
	&&+ { g^2\over 4 m^2} \,\left.
	 j^ {\mu\nu}\, { 1\over \Box} \, j_ {\mu\nu}\,\right]
        \nonumber\\
	 = \int d^4x  &&\,\left[\,
        {1\over 2}\, \left(\, \Lambda - g^2\,\Theta(x)\,\right)
         -{m^2\over 2\cdot 3!}\,
        \widehat A^{\mu\nu\rho}\, \left(\, {\Box +  m^2\over \Box}\,\right)\,
	\widehat A_{\mu\nu\rho} + {g\,m^2\over 3!}\,
	\widehat J^{\mu\nu\rho}\, {1 \over\Box}\, \widehat
	A_{\mu\nu\rho}\right.\nonumber\\
	&&+{ g^2\over 4 m^2 }  \left. j^ {\mu\nu}\, { 1\over \Box }\,
	j_ {\mu\nu}\,\right]\ .
        \label{stens}
        \end{eqnarray}

     	The first term is of the same form as in (\ref{13}).
     	It represents the bubble ``volume action'' with respect to the
constant energy
     	density background that determines the false vacuum decay rate. \\
The second and third term govern the dynamics of $\widehat A_{\mu\nu\rho}$
according to the equation
     	\begin{equation}
	 \left(\, \Box +  m^2\, \right)\, \widehat  A_{\mu\nu\rho} = g\,
	{ \widehat J^{\mu\nu\rho} }\ .
	\end{equation}
We emphasize that this massive mode represents the {\it only} propagating
degree of freedom, exactly as in the $(1+1)$ dimensional case. As a matter
of fact, the last term in the action
(\ref{stens}) represents a {\it boundary induced Coulomb interaction}
\cite{luscher}.
Indeed, a direct calculation taking into account $\partial_\mu \,j^{\mu\nu}=0$
 gives

     	\begin{eqnarray}
     	{ g^2\over 4 m^2 }\int d^4x\, j^ {\mu\nu}\, { 1\over \Box }\,
     	 j_ {\mu\nu}&=&
     	{ g^2\over 2 m^2 } \int dx^0\, d^3x\, j^ {0 k}(x^0,\vec x )\,
     	 { 1\over \nabla^2}\, j^{0k}(x^0,\vec x )\nonumber\\
     	 &=&
     	{ g^2\over 2 m^2 } \int dx^0 \int d^3x \int d^3y\,
     	 j^ {0 k}(x^0,\vec x )\, { 1\over \nabla^2}\,
     	 \delta^{3)}\left(\,\vec x -\vec y \,
     	 \right)\, j^{0k}(x^0,\vec y )\nonumber\\
     	 &=&
     	-{ g^2\over 4\pi m^2 } \int dx^0 \int d^3x \int d^3y\,
     	j^ {0 k}(x^0,\vec x )\,
     	{ 1\over |\,\vec x -\vec y\,|} \, j^{0k}(x^0,\vec y )
     	 \end{eqnarray}

showing that there is no physical particle mediating such an interaction.

\subsection{Restoring Gauge Invariance }

        In the previous subsections we have developed a self--consistent
        model for membranes with a spacelike boundary, coupled to a
        massive tensor field. The price for that result is the apparent loss of
manifest gauge invariance. As in the case of ``bubble dynamics'' in $(1+1)$
dimensions, this leads us to the question: is there a way of
introducing a mass term into the action without spoiling manifest gauge
invariance?
Once again, we follow the original Stueckelberg proposal\cite{stuck} of
restoring gauge invariance by introducing a mass term together with a
compensating scalar field. Presently, however, we need a modification of
Stueckelberg's approach that is suitable for our massive tensor theory of
bubble dynamics. The procedure is
        straightforward. The only novel aspect is that the role of
compensating field is now played by a two--index Kalb--Ramond potential
        $ B_{\nu\rho}(x)$\cite{kr}. Accordingly, we modify the action
	(\ref{massmem}) as follows:

	\begin{eqnarray}
        S&=&\int d^4x\left[\, {1\over 2\cdot 4!}\, F_{\lambda\mu\nu\rho}\,
        F^{\lambda\mu\nu\rho}-{1\over  4!}\, F^{\lambda\mu\nu\rho}\,
        \partial_ {\,[\,\lambda}A_ {\mu\nu\rho\,]}
        -{g\over 3!}\, J^{\mu\nu\rho}\, \left(\,
        A_{\mu\nu\rho}+{1\over m}\,\partial_{\,[\,\mu}B_{\nu\rho\,]}\,\right)
        \right.\nonumber\\
        &&-{m^2\over 2\cdot 3!}\left. \left(\,
	 A_{\mu\nu\rho}+{1\over m}\,\partial_{\,[\,\mu}B_{\nu\rho\,]}\,\right)
	\left(\,
         A^{\mu\nu\rho}+{1\over m}\, \partial^{\,[\,\mu}B^{\nu\rho\,]}\,\right)
         \,\right]
	\label{lopen}
        \end{eqnarray}
	This action is invariant under the extended tensor gauge transformation
	\begin{eqnarray}
	&&\delta A_{\mu\nu\rho}=\partial_ {[\,\mu}\Lambda_ {\nu\rho\,]}
	\label{simmuno}\\
	&&\delta B_{\nu\rho}= -m \,\Lambda_ {\nu\rho}\ .
\label{simmdue}
	\end{eqnarray}

	Note that the (gauge invariant) kinetic term for $B$ makes $A$
        massive: from a dynamical point of view, the presence of a boundary,
        or a non conserved current, introduces a mass term for the
        gauge field that the current is coupled to.\\
        The field equations become
	\begin{eqnarray}
        &&\partial_\lambda\, F^{\lambda\mu\nu\rho}+m^2\,
        \left(\, A^{\mu\nu\rho} +{1\over m}\,
        \partial^{\,[\,\mu}B^{\nu\rho\,]}\,
	\right)=g\, J^{\mu\nu\rho}
	\label{uno}\\
        &&\partial_\mu\, \left(\, A^{\mu\nu\rho}
	+{1\over m}\, \partial^{\,[\,\mu}B^{\nu\rho\,]}\, \right)
        ={g\over m^2}\, J^{\nu\rho}
        \label{due}
        \end{eqnarray}
	Equation (\ref{due}) assures the self--consistency of Eq.
       (\ref{uno}). Moreover, using the same field decomposition as in the
        previous section, we see that equation (\ref{due}) fixes the
        divergenceless component $\widetilde A_{\mu\nu\rho} $:

	\begin{equation}
	\widetilde A^{\mu\nu\rho} + {1\over m }\, \partial^{\,[\,\mu}
	B^ {\nu\rho\,]}={e\over m^2}\,
	\partial^{\,[\, \mu }\,{1\over \Box }\,  J^{\nu\rho\,]}
	\label{longim}
	\end{equation}

	while, Eq.(\ref{uno}) leads to the following expression for
        $F^{\lambda\mu\nu\rho}$:

	\begin{eqnarray}
	F^{\mu\nu\rho\sigma}&=& \sqrt\Lambda\, \epsilon^{\mu\nu\rho\sigma}\,
	+ g^2\,  \partial^{\, [\,\mu }\,{1\over \Box}\,
	\widehat J^{\nu\rho\sigma\,]}
	 -m^2\,  \partial^{\, [\,\mu}\,{1\over \Box}\,
	 \widehat A^{\nu\rho\sigma\,]}
	\nonumber\\
	&\equiv & F^{\mu\nu\rho\sigma}_0 -m^2\,
	\partial^{\, [\,\mu}\,{1\over\Box}\,{\widehat A}^{\nu\rho\sigma\,]}\ .
	\end{eqnarray}

         Substituting the above expression into the action, we obtain after
         some rearrangement

        \begin{eqnarray}
        S &=&  \int d^4x\left[\, {1\over 2\cdot 4! }\,
        F_ {0\,\,\mu\nu\rho\sigma}\, F_ 0^{\mu\nu\rho\sigma}
        -{m^2\over 2\cdot 3!}\,
        \widehat A^{\mu\nu\rho}\,\left(\, {\Box +  m^2\over \Box}\,\right)
	\widehat A_{\mu\nu\rho}+ \right.\nonumber\\
	 &+& {g\,m^2\over 3!}\,
	\widehat J^{\mu\nu\rho}\, {1 \over\Box}\, \widehat A_{\mu\nu\rho}
	+\left.  { g^2\over 4 m^2} \, j^ {\mu\nu}\, { 1\over \Box}\, j_
{\mu\nu}
        \,\right]\label{esseqfine}  \\
	  &=& - { 1\over 2}\, \int d^4x\left[\,\Lambda +{g^2\over 2\cdot 3!}\,
	\widehat J^{\mu\nu\rho}\,  { 1\over \Box +  m^2 }\,
	\widehat J_ {\mu\nu\rho} -{ g^2\over 4 m^2}\,
	j^ {\mu\nu} \,  {1\over\Box }\, j^ {\mu\nu}  \,\right]
	\label{essefine}
	 \end{eqnarray}

	The final form of the action (\ref{essefine}) shows how
	the introduction of a compensating Kalb--Ramond field, which is
	necessary for restoring gauge invariance, leads to an
	{\it ``effective closure''} of the membrane in the physical Minkowskian
	spacetime and is, in fact, an alternative to the euclidean procedure of
	 closing the membrane in imaginary time.\\
        However, it seems to us that a careful consideration of the boundary
        effect in the nucleation process of a vacuum bubble in real spacetime
        has a clear advantage over the euclidean formulation in that it
brings out the existence of a massive pseudo--scalar degree of freedom which
is otherwise hidden in the energy background provided by the cosmological
field.

\section{Conclusions and Outlook}
\label{fine}

To the extent that the presence or absence of a boundary constitutes a
topological property of a manifold, we may refer to the idea underlying the
whole discussion in this paper as {\it topological symmetry breaking.} The
effect of this new mechanism on the inflationary--axion scenario is
apparent in our formulation of Bubble Dynamics, and was illustrated in four
as well as in two spacetime dimensions. Indeed, the action functional of
Bubble Dynamics
can be defined in any number of dimensions as a generalization of the
Einstein--Maxwell action for the dynamics of point charges on a Riemannian
manifold. As a matter of fact, gravity plays no special role in it: even
though we have formulated the model for a bubble in $(3+1)$ Minkowski
spacetime, it can be extended to a generic $p$--brane embedded in a
target space with $D=p+2$ dimensions. In four dimensions and under the
assumption of spherical symmetry, the field equations of bubble dynamics are
integrable\cite{acl}: the net physical result of the $A_{\mu\nu\rho}$ coupling
to the membrane degree of freedom is the nucleation of a bubble whose boundary
separates two vacuum phases characterized by two effective and distinct
cosmological constants, one inside and one outside the bubble \footnote{
The evolution of the bubble,
which is controlled by the two cosmological constants and by the surface
tension, can be simulated  by the one--dimensional motion of a fictitious
particle in a potential\cite{noi4}; furthermore, a well defined algorithm
exists that is capable of determining all possible types of solutions,
including inflationary ones, together with the region in parameter space where
families of solutions can exist\cite{noi5}}.\\
The three--index representation of the cosmological field is the key to the
whole formulation of topological symmetry breaking and self--consistent
generation of mass. That field, we have argued, represents
the ultimate source of energy in the bubble universe. But how does matter
manage to `` bootstrap '' itself into existence out of that source of latent
energy? Here is where the difference between the conventional
``cosmological constant'' and the cosmological field comes into play: the
original cosmological constant introduced by Einstein plays a somewhat
passive role, in that it is ``frozen'' within the Hilbert action of general
relativity; the cosmological field, on the contrary, even though it
represents a non dynamical gauge field, will {\it interact} with gravity
and combine with a bubble single degree of freedom thereby acquiring mass
as a consequence of topological symmetry breaking. In fact, there are some
similarities with the Higgs mechanism which may help to clarify the
boundary mechanism of mass production. For instance, the extended
gauge symmetry (\ref{simmuno}), (\ref{simmdue}) represents the end result
of a process that begins with the violation of gauge invariance
due to the presence of a boundary. Then,
the Kalb--Ramond field $B_{\mu\nu}$, prescribed by the Stueckelberg
procedure, represents massless, spinless particles that play the role of
Goldstone bosons. However, while in the usual Higgs
mechanism the spin content of the gauge field is the same before and after
the appearance of mass, a new effect occurs when the gauge field is
$A_{\mu\nu\rho}$: when massless, $A_{\mu\nu\rho}$
carries no degrees of freedom, while equation (\ref{unobis})
describes massive spin--0 particles in a
representation which is {\it dual} to the familiar Proca representation of
massive, spin--1 particles\cite{au}. In the light of this formal analogy with
the Higgs mechanism, axions are interpreted as massless spin--0 Goldstone
bosons represented by a Kalb--Ramond field while the physical spectrum
consists of massive spin--0 particles represented by the $\widehat
A_{\mu\nu\rho}$--field. In this sense, topological symmetry breaking by the
boundary has the same effect as the breaking of the Peccei--Quinn symmetry
in the {\it local} standard model of particle physics.\\
\\
We do not have at present a fully fledged quantum theory of
bubble--dynamics even though we have taken several steps on the way to that
formulation\cite{progress},\cite{noi10}. However, if bubble dynamics in two
dimensions is any guide, one can anticipate the main features of the
quantum theory: as the volume of the bubble universe
increases exponentially during the inflationary phase, so does the total
(volume) energy of the interior ``De Sitter vacuum''. At least classically.
Quantum mechanically there is a competitive effect which is best understood
in terms of an analogous effect in (1+1)--dimensions. As we have argued in
the previous section, the``volume'' within a one--dimensional bubble is the
linear distance between the two end--point charges. As
the distance increases, so does the potential energy between them.
Quantum mechanically, however, it is energetically more favorable to
polarize the vacuum through the process of pair creation \cite{kog}, which we
interpret as the nucleation of secondary bubbles out of the vacuum enclosed
by the original bubble. The net physical result of this mechanism is the
production of massive spin--0 particles \cite{schw}. The same mechanism can be
lifted to $(3+1)$--dimensions and reinterpreted in the cosmological context:
the $A_{\mu\nu\rho}$ field shares the same properties of the gauge potential
$A_\mu$ in two dimensions and polarizes the vacuum via the formation of
secondary bubbles.
Consider now a spherical bubble and focus on the radial evolution alone.
The intersection of any diameter with the bubble surface evolves precisely
as a particle--anti particle pair in $(1+1)$--dimensions. However, since there
is no preferred direction, the mechanism operates on concentric {\it shells}
inside the original bubble. Remarkably, the final result is again the
production of massive pseudoscalar particles in the bubble universe.
However, while in two dimensions Goldstone bosons do not exist, in
$(3+1)$--dimensions they do exist and have a direct bearing on the axion
mass problem. Born out of the darkness of the cosmic vacuum, axions were
invisible to begin with and remain invisible to the extent that they are
`` eaten up '' by the cosmological field. According to this interpretation,
one of the possible forms of dark matter in the universe, besides MACHO's,
emerges as the
{\it necessary} end product of a process, driven by the cosmic vacuum
energy, according to which
gauge invariance and vacuum decay conspire to extract massive particles
out of the cosmological field of dark energy.

\end{document}